\documentclass[prl,preprintnumbers,twocolumn]{revtex4} 

\usepackage{graphicx}
\usepackage{dcolumn}
\usepackage{bm}
\usepackage{latexsym}
\usepackage{amsfonts}
\usepackage{amssymb}
\usepackage{amsmath}
\usepackage{mathrsfs}
\usepackage{bbold}
\usepackage{ulem}
\usepackage{url}
\usepackage[utf8]{inputenc} 
\usepackage{xspace}

\begin{document}

\title{Role of matter in gravitation: going beyond the Einstein-Maxwell theory}
\author{A. Emir G\"umr\"uk\c{c}\"uo\u{g}lu}
\email{emir.gumrukcuoglu@port.ac.uk}
\affiliation{Institute of Cosmology and Gravitation, University of Portsmouth\\ Dennis Sciama
Building, Portsmouth PO1 3FX, United Kingdom}

\author{Ryo Namba}
\email{namba@physics.mcgill.ca}
\affiliation{Department of Physics, McGill University, Montr\'eal, QC, H3A 2T8, Canada}
\date{\today}

\begin{abstract}
For field theories in curved spacetime, defining how matter gravitates is part of the theory building process. In this letter, we adopt Bekenstein's multiple geometries  approach to allow part of the matter sector to follow the geodesics on a general pseudo-Riemannian geometry, constructed from a tensor and a $U(1)$ gauge field. This procedure allows us to generate a previously unknown corner of vector-tensor theories. In the Jordan frame, apparent high-derivative terms of the vector field are reduced by integrating out an auxiliary variable, at the cost of introducing new matter interactions.
As a simple example, we consider a conformal relation between different geometries and demonstrate the presence of an auxiliary degree. We conclude with a discussion of applications, in particular for the early universe.
\end{abstract}

\maketitle

\section{Introduction}

Extensions of Maxwell's electrodynamics, the linear theory of $U(1)$ gauge fields, have been actively studied in the last century. One of the most well known examples is the nonlinear construction by Born and Infeld \cite{Born:1934gh}. This theory, which was designed to address the divergences in the self-energy of point charges, can be considered as a special case of a class of theories where the Lagrangian is an arbitrary function of the Maxwell term $F_{\mu\nu}F^{\mu\nu}$ and the Pontryagin density $F_{\mu\nu}\tilde{F}^{\mu\nu}$. In the absence of gravity, this class is the most general classical action for a $U(1)$ field, as indicated by the no-go result in Ref.~\cite{Deffayet:2013tca}.

On the other hand, in curved spacetime, one can devise highly nonlinear interactions between the gauge field and gravity. A known extension of the Einstein-Maxwell theory to include such terms has been constructed by Horndeski \cite{Horndeski:1976gi}, requiring that the equations of motion are at most of second order. This requirement is justified by Ostrogradski's result \cite{Ostrogradsky:1850fid} which demonstrates that the Hamiltonian becomes unbounded from below if the equations of motion contain more than second-order time derivatives. However, with degenerate kinetic terms, it is possible to evade this conclusion; the implicit constraints due to the degeneracy render a potentially problematic degree of freedom auxiliary \cite{Langlois:2015cwa}.

In the present paper we exploit this loophole to find new extensions of $U(1)$--tensor theory. Instead of building a general class from first principles, we prove the presence of these theories by adopting a perspective based on Bekenstein's multiple geometries \cite{Bekenstein:1992pj}. 
In this approach, one starts from a gravitational action in vacuum, then adds matter that follows the geodesics of an arbitrary geometry built out of all degrees of freedom that participate in gravitational interactions. An essential advantage of this strategy is that a second-order equation of motion in one representation can seem to be of higher order in a different (but equivalent) representation, at the expense of an auxiliary degree of freedom. This convenient method has previously been used, at least partially, in the context of scalar-tensor \cite{Bettoni:2013diz,Zumalacarregui:2013pma} and massive vector-tensor theories \cite{Jimenez:2016isa,Kimura:2016rzw,Domenech:2018vqj}. To our knowledge, the present study is the first attempt in which the approach is used to generate new gravitational interactions of a $U(1)$ gauge field.

In our construction, the relation between the two geometries is chosen to preserve the form of the local $U(1)$ symmetry for a vector field $A_\mu \to A_\mu + \partial_\mu \alpha$ in both representations. The novel interactions between this gauge field $A_\mu$ and the metric potentially provide several implications especially for the physics of the early universe, such as inflationary dynamics, generation of magnetic fields, and spacetime singularities. 
Since the new interactions are invoked by the relation between the Jordan and Einstein frames by construction, they naturally become ineffective in the region of low matter density and vanish in the absence of matter. In the context of cosmology, this potentially allows our theory to generate primordial signatures without violating current local experiments.

\section{A tale of two geometries}

When the $U(1)$ gauge field is identified as a gravitational degree of freedom, the metric ceases to be the unique quantity that characterizes the geometry. In fact, starting from an arbitrary metric variable $g$, we can define a new one through
\begin{equation}
\tilde{g}_{\mu\nu} = C\,g_{\mu\nu} +D\, F_{\mu\rho} g^{\rho\sigma}F_{\sigma\nu}\,,
\label{eq:disformal}
\end{equation}
where $F_{\mu\nu}=\partial_\mu A_\nu-\partial_\nu A_\mu$ is the field strength tensor, while the coefficients $C$ and $D$ vary with both the tensor and vector field
\begin{equation}
C=C([F^2],[F^4])\,, \qquad D=D([F^2],[F^4])\,,
\end{equation}
with square brackets denoting the trace operation $[F^2] = F^{\mu}_{\;\;\nu} F^{\nu}_{\;\;\mu}$ and $[F^4] = F^{\mu}_{\;\;\nu} F^{\nu}_{\;\;\rho}F^{\rho}_{\;\;\sigma}F^{\sigma}_{\;\;\mu}$. The first term in \eqref{eq:disformal} corresponds to a conformal transformation, i.e. isotropic stretching of spacetime distances. The second term is the only $U(1)$ invariant disformal (anisotropic stretching) contribution that is \textit{i.} compatible with the symmetry of the metric tensor; \textit{ii.} contains no more than first derivatives of the vector field. 
Note that, although one can add other disformal terms containing any even powers of $F^\mu_{\;\;\nu}$, they can be absorbed in the definitions of the coefficients $C$ and $D$ thanks to the Cayley-Hamilton theorem in $4$ dimensions.
Equation~\eqref{eq:disformal} is thus the most general relation between two metrics that requires only local information of field values (zeroth and first derivatives), respecting the $U(1)$ symmetry.
As we shall see, it is also the most general one, barring a fine-tuned choice of $C$ and $D$, that does not introduce additional degrees of freedom.

By changing the matter coupling according to relation \eqref{eq:disformal}, one can generate new vector-tensor theories. For instance, starting from a vector-tensor Lagrangian $\mathcal{L}_{\rm VT}$ whose variation yields second order equations of motion, one can introduce the matter sector that no longer follows the geodesics of the tensor in the original Lagrangian, but a metric that is disformally related to it, which can be characterized by the form of action
\begin{equation}
S = \int d^4x \left(\sqrt{-\tilde{g}} \,\mathcal{L}_{\rm VT} ( \tilde{g},A_\mu)+ \sqrt{-g}\, \mathcal{L}_{\rm m}(g,\{\Psi_I\}) \right)\,,
\label{theory}
\end{equation}
where $\{\Psi_I\}$ represents a set of matter fields. Provided that the matter Lagrangian does not contain any more than first derivatives of the fields, the resulting equations of motion for the $\tilde{g}$ representation are guaranteed to have manifestly second derivatives. On the other hand, when one instead chooses the $g$ representation, which would be the Jordan frame, a curvature dependent $\mathcal{L}_{\rm VT}$ leads to terms in the equations of motion that have higher than second derivatives. 

For instance, the Einstein-Hilbert term $S_{\rm EH} = \int d^4x \sqrt{- \tilde{g}} \, \tilde{R}$ in the $\tilde{g}$ representation can be written (up to boundary terms) in the $g$ representation as
\begin{widetext}
\begin{align}
S_{\rm EH} & 
=  \int d^4x 
\sqrt{-g} \, C \,{\cal E} \,\Bigg[ \mathfrak{h}^{\mu\nu} \left( R_{\mu\nu} 
+ \frac{3}{2} \, \frac{\nabla_\mu C \, \nabla_\nu C}{C^2} 
+ 2 \, \frac{\nabla_\mu C \, \nabla_\nu {\cal E}}{C {\cal E}} 
+ \frac{\nabla_\mu {\cal E} \, \nabla_\nu {\cal E}}{{\cal E}^2}
+ \frac{1}{2} \nabla_\mu \mathfrak{h}^{\rho\sigma} \left( \frac{1}{2} \nabla_\nu h_{\rho\sigma} - \nabla_\sigma h_{\nu\rho} \right) \right) 
\nonumber\\ & \qquad\qquad\qquad\qquad\quad
+ \nabla_\mu \mathfrak{h}^{\mu\nu} \left( \frac{\nabla_\nu C}{C} + \frac{\nabla_\nu {\cal E}}{{\cal E}} \right)
\Bigg] \; ,
\label{SEH}
\end{align}
\end{widetext}
where $h_{\mu\nu} \equiv g_{\mu\nu} + \frac{D}{C} F_\mu{}^\rho F_{\rho\nu}$, $\mathfrak{h} \equiv h^{-1}$ and ${\cal E}^2 \equiv \det (g^{-1}h)$. Since $C$, $\mathcal{E}$, $h_{\mu\nu}$ and $\mathfrak{h}_{\mu\nu}$ all depend on the first derivative of the vector field, the action above generically contains terms of the form $(\partial\partial A)^2$. However, provided that the two representations are equivalent, the high derivatives do not lead to an Ostrogradski-type instability; instead the apparent new degree of freedom implied by them is an auxiliary field. 

The equivalence of the two representations is only true if the disformal relation \eqref{eq:disformal} is invertible.
Invertibility ensures that no information is lost after a change of variables. The theory thus continues to be invariant under general coordinate transformations and the Ostrogradski stability arguments in the previous representation continue to hold. In order to obtain the conditions for invertibility, we need the Jacobian of the two metrics
\begin{equation}
J^{\alpha\beta}_{\;\;\;\;\mu\nu} \equiv \frac{\partial \tilde{g}_{\mu\nu}}{\partial g_{\alpha\beta}}\,.
\end{equation}
The eigenvalues of the Jacobian can be calculated by following the general prescription laid out in Ref.~\cite{Zumalacarregui:2013pma}. Requiring no null eigenvalues, we find the invertibility condition as
\begin{widetext}
\begin{align}
& C\,\left[ C-2\,C_2 [F^2]-4\,C_4 [F^4]- \frac{D\,[F^2]}{2}-2\,D_2[F^4]+\frac{D_4[F^2]\,([F^2]^2-6\,[F^4])}{2} \right]
\nonumber\\ & \quad
+ \left( [F^2]^2-2\,[F^4] \right)
\left[ \frac{D}{8} \left(D+2\,D_2[F^2]+4\,D_4[F^4]+8\,C_2+4\,C_4[F^2]\right) 
+ \left( C_4D_2-C_2D_4 \right) \left([F^2]^2-4\,[F^4]\right)
\right]
\neq0\,,
\label{eq:invertibility}
\end{align}
\end{widetext}
where we have defined
\begin{equation}
C_n \equiv \frac{\partial C}{\partial [F^n]}\,,\qquad
D_n \equiv \frac{\partial D}{\partial [F^n]}\,.
\end{equation}

In the following, we turn off the disformal term and focus on a conformal relation between two representations. Although this is a highly simplified case, it provides an illustrative example that demonstrates the properties discussed above.

\section{Conformally related geometries}
As an example, we consider the Einstein-Maxwell theory, then couple the rest of the matter sector to a new metric
\begin{equation}
S =  \int d^4x \left[ \sqrt{-\tilde{g}}\left(
\frac{\tilde{g}^{\mu\alpha}\tilde{g}^{\nu\beta}}{4}F_{\mu\nu}F_{\beta\alpha}+\frac{M^2}{2} \tilde{R}
\right) +\sqrt{-g} {\cal L}_{\rm m} \right] \,,
\label{eq:action}
\end{equation}
where $g$ and $\tilde{g}$ are now related only conformally, i.e.~$D=0$, and $M$ is a constant of mass dimension $1$ that would be identified as the reduced Planck mass in the limit $C \to 1$. In this case, the invertibility condition \eqref{eq:invertibility} reduces to
\begin{equation}
C\,\left(C-2\,C_2 [F^2]-4\,C_4 [F^4]\right) \neq 0\,.
\end{equation}
 When written in the Jordan frame using the $g$ metric as the variable, the action has the familiar form of a conformally transformed system, up to surface terms,
\begin{align}
\!\!
S=& \int d^4x\sqrt{-g}\left[ 
 \frac{M^2 C}{2}\left(R+\frac{3}{2}\, \frac{\nabla^\mu C \, \nabla_\mu C}{C^2} \right) 
 +\frac{[F^2]}{4} + \mathcal{L}_\text{m}
 \right]\,.
\label{action_conf}
\end{align}
As discussed earlier, the derivatives of the conformal factor generate terms of the form $(\partial\partial A)^2$, which generically lead to equations of motion of fourth order in derivatives. Varying the Jordan frame action \eqref{action_conf} with respect to the gauge field and the metric leads to, respectively,
\begin{widetext}
\begin{align}
&\nabla_\nu \mathcal{F}^{\mu\nu} = 0 \; , \qquad\qquad
\mathcal{F}^{\mu\nu} \equiv 
F^{\mu\nu} + 
2 M^2 C \left( R
- 3 \, \frac{\nabla^2 C}{C} 
+ \frac{3}{2} \, \frac{\nabla^\rho C \, \nabla_\rho C}{C^2} \right) 
\left( \frac{C_2}{C} F^{\mu\nu} + 2 \, \frac{C_4}{C} \left( F^3 \right)^{\mu\nu} \right) 
 \; ,
\label{EOM_gauge}
\\
& G_{\mu\nu} 
+ \frac{3}{2} \, \frac{\nabla_\mu C \, \nabla_\nu C}{C^2}
- \frac{\nabla_\mu \nabla_\nu C}{C} 
+ \left( \frac{\nabla^2 C}{C}
- \frac{3}{4} \, \frac{\nabla^\rho C \nabla_\rho C}{C^2} \right) g_{\mu\nu}
=
\frac{1}{M^2 C}\left( T_{\mu\nu} 
- \mathcal{F}_{\mu\rho} F^\rho{}_\nu
+ \frac{\left[ F^2 \right]}{4} \, g_{\mu\nu} \right) \; .
\label{modifiedEinstein}
\end{align}
\end{widetext}
Indeed, the vector field equations \eqref{EOM_gauge} contain up to fourth derivatives of $A_\mu$, while the metric field equations \eqref{modifiedEinstein} go up to third order. However, the auxiliary degree of freedom can be integrated out by taking the trace of Eq.~\eqref{modifiedEinstein},
\begin{align}
&\!\!\!\!\!\!\!\!\!\!\left( R
- 3 \, \frac{\nabla^2 C}{C} 
+ \frac{3}{2} \, \frac{\nabla^\rho C \nabla_\rho C}{C^2} \right) 
\left( \frac{C_2}{C} \left[ F^2 \right]
+ 2 \, \frac{C_4}{C} \left[ F^4 \right]
- \frac{1}{2}
\right)
\nonumber\\
&\qquad\qquad\qquad\qquad\qquad\qquad\qquad\qquad= \frac{T}{2 M^2 C} \; .
\label{eq:constraint}
\end{align}
The first parenthesis contains the only third derivative term which also appears in the definition of $\mathcal{F}_{\mu\nu}$ in \eqref{EOM_gauge}. Using Eq.~\eqref{eq:constraint} we can thus reduce the derivative order of the vector equation down to two. Rewriting Eq.~\eqref{EOM_gauge}, we get
\begin{equation}
\nabla_\nu 
 \left[
F^{\mu\nu} + 
\left(
\frac{\frac{C_2}{C} F^{\mu\nu} + 2 \, \frac{C_4}{C} \left( F^3 \right)^{\mu\nu} 
}{
 \frac{C_2}{C} \left[ F^2 \right]
+ 2 \, \frac{C_4}{C} \left[ F^4 \right]
- \frac{1}{2}
}\right)\,T
 \right] =0\,.
\label{EOM_gauge2}
\end{equation}
This equation is now manifestly second order, and part of the metric equations \eqref{modifiedEinstein} that corresponds to the Hamiltonian constraint also reduces to second (time) derivatives of $A_\mu$, closing the system of equations that govern time evolution. Removing the auxiliary degree introduces an explicit dependence on the trace of the energy-momentum tensor in the vector equation. In the Jordan frame, the nontrivial metric--gauge field interaction thus implies a direct coupling between the matter and the gauge field, as was already manifest in the Einstein frame. We remark that for a matter action that does not contain first derivatives of the metric, the matter energy-momentum tensor is covariantly conserved and the matter equation of motion is oblivious to the constraint \eqref{eq:constraint}. Therefore, standard fermionic fields have their usual equation of motion in the Jordan frame.

The reduction of higher derivatives is possible only in the presence of the metric degrees of freedom. If one turned off gravity at the level of the action \eqref{action_conf}, the metric equation \eqref{modifiedEinstein} and thus its trace \eqref{eq:constraint} would be absent, removing essential geometrical information to eliminate the higher derivatives in the gauge-field equation of motion \eqref{EOM_gauge}. This observation suggests that the spacetime geometry cannot be treated as external, and the dynamics of the metric is crucial to ensure the absence of higher-derivative instabilities.

This construction is especially interesting in the context of the early universe cosmology. The conformal invariance of the standard Maxwell term in $4$ dimensions leads to an effective decoupling of the gauge field from expansion, posing a challenge for simplest models of generation of primordial magnetic fields \cite{Ratra:1991bn}. The matter-gauge coupling in \eqref{EOM_gauge2}, or equivalently the high-derivative degenerate terms in \eqref{EOM_gauge}, can potentially allow the vector field to be susceptible to the expansion. 

The matter sector is dominated by the inflaton field during inflation. For a perturbative gauge field production, the equation of motion at leading order is 
\begin{equation}
\nabla_\nu 
 \left[
F^{\mu\nu} \left(1-2\,\frac{C_2}{C}\,T \right)
 \right] =0 \,,
\label{EOM_linear}
\end{equation}
where $C_2/C$ is evaluated at $[F^2] = [F^4] = 0$ and is constant. We require $2 T C_2 / C < 1$ to avoid ghost instabilities, and once $T$ sufficiently decreases, Eq.~\eqref{EOM_linear} returns to the standard Maxwell equation without a source.
Eq.~\eqref{EOM_linear} is equivalent to a model with Lagrangian density $f(t) F_{\mu\nu}F^{\mu\nu}$, where the time dependence of function $f(t)$ is induced by the motion of the inflaton condensate. This model is known to suffer difficulty to generate sufficiently large magnetic fields to account for the blazar observations \cite{Neronov:1900zz} (see \cite{Durrer:2013pga} for a review). This is because under the requirement to avoid a strong coupling to charged particles, electric fields are always produced at larger amplitudes than magnetic. Imposing bounds on the background dynamics as well as curvature perturbation constraints in turn suppresses the level of magnetic-field production. The linear production from \eqref{EOM_linear} falls into the same class, and this situation would not change even if the disformal factor $D$ in \eqref{eq:disformal} is included.

Another possibility is non-linear generation of a magnetic field condensate as an attractor solution, in the fashion of Ref.~\cite{Mukohyama:2016npi}. With the disformal term turned off, however, this scenario would not be successful either, due to the lack of a scaling symmetry that could ensure that the magnetic field persists against the expansion of spatial volume.

\section{Discussion}
In this paper, we proposed a novel extension of vector-tensor theories, where the vector field is an Abelian gauge field. We introduced a previously overlooked disformal relation between two geometries as a method to construct this theory. To demonstrate the strength of our approach, we considered a simple conformal relation, which is sufficient to generate a new theory with higher-order derivatives that is immune to Ostrogradski instabilities thanks to the presence of an auxiliary field. 

Non-linear extensions of the Einstein-Maxwell theory provide a novel, intriguing framework to address problems in the early universe, such as new inflationary solutions, primordial magnetogenesis and avoidance of initial singularities. Other interesting applications include problems in the strong gravity regime; for instance, determining the implications of the nonlinear gauge field--tensor interaction for compact object solutions.
We will present a more complete discussion of magnetogenesis, black hole solutions and spacetime singularity avoidance for the most general disformal transformation in a future publication.
A key ingredient in our construction of the theory \eqref{theory} relevant to address these issues is that gravitational interaction is mediated not only by the metric but also by an additional gauge field, potentially changing the solutions in a way otherwise impossible, and that such effects arise particularly in large density regimes.

A potential concern is that this construction breaks one or more equivalence principles. Since in the Jordan frame the gravitational coupling gets redressed by the vector-field dependent terms, this is a manifestation of strong equivalence principle violation. Moreover, if one considers the gauge field to be the electromagnetic field, it will have direct coupling to the Standard Model fields that do not even carry any $U(1)$ charge. Although there is no single resolution that categorically implies compatibility with observations and experiments, one can construct realistic scenarios with reasonable implications depending on the context. For instance, there are several new scales that are introduced by the relation \eqref{eq:disformal}, such as $(C_2)^{-1/4}$ or $(C_4)^{-1/8}$, which can be tuned to suppress nonstandard gravitational effects.

Another plausible scenario to overcome the above concern is to preserve strong equivalence principle for the gravitational interactions of the Standard Model sector, but violate it for a sector of new physics. For example, one can violate the weak equivalence between the inflaton sector and Standard model, by devising a model with the action
\begin{align}
S =& \int d^4x \left[\frac{M^2}{2}\sqrt{-g} \, R - \frac14\sqrt{-g}\, F_{\mu\nu}F^{\mu\nu}
+ \sqrt{-\bar{g}}\, \mathcal{L}_{\rm inf}(\bar{g},\phi) 
\right.\nonumber\\
&\qquad\quad
\left.
+\sqrt{-g}\, \mathcal{L}_{\rm SM}(g,\{\Psi_I\}) \right]\,,
\label{eq:wep-violation-scenario}
\end{align}
with $\bar{g}$ and $g$ being related by a disformal relation of the form \eqref{eq:disformal}. 
In this case, inflaton $\phi$ and the Standard Model fields follow the geodesics of different geometries. The Jordan frame from the perspective of the Standard Model coincides with the Einstein frame. After the inflaton decays, the vector field becomes a standard Maxwell field in curved background. Keeping the metric in the Jordan frame of Standard Model, high derivatives of the vector fields do not emerge. In this form, the scalar field can be interpreted as part of the gravity sector, now consisting of a scalar, vector and tensor fields, as proposed previously in \cite{Heisenberg:2018acv}.
In this case, one can identify the $U(1)$ field with the photon without causing any inconsistency, since the weak equivalence principle is reinstated in the post-inflationary stage. This type of implementation is quite relevant for early universe problems, since the non-linear behaviour is constrained to the inflationary stage. 

The relation \eqref{eq:disformal} can also be used to further extend massive vector--tensor theories proposed in Refs.~\cite{Heisenberg:2014rta,Hull:2015uwa}. In these theories, $U(1)$ symmetry is absent, and thus the vector field has a third (longitudinal) polarization. For this reason, previous applications of disformal transformations did not include dependence on the electromagnetic strength tensor. On the other hand, since the disformal term $F_{\mu\rho}F^{\rho}_{\;\;\nu}$ in \eqref{eq:disformal} is invariant under $U(1)$ transformation, we do not expect it to excite non-degenerate high-derivative ghosts associated with the longitudinal mode in any representation.

Finally, throughout the paper, we assumed parity invariance. However, the approach can be extended to include dependence on the dual tensor $\tilde{F}^{\mu\nu}$, which amounts only to making $C$ and $D$ depend on a pseudo-scalar quantity $F_{\mu\nu} \tilde{F}^{\mu\nu}$ in addition to $[F^2]$ and $[F^4]$.

\vspace{.5cm}
\begin{acknowledgments}
We thank Tomohiro Fujita, Kazuya Koyama, Shinji Mukohyama, Alexander Vikman and Daisuke Yoshida for illuminating discussions.  The work of AEG has received funding from the European Research Council (ERC) under the European Union’s Horizon 2020 research and innovation programme (grant agreement No.~646702 ``CosTesGrav'').
The research of RN is in part supported by the Natural Sciences and Engineering Research Council (NSERC) of Canada
and by the Lorne Trottier Chair in Astrophysics and Cosmology at McGill University.

\end{acknowledgments}



\begin{thebibliography}{99}

\bibitem{Born:1934gh} 
  M.~Born and L.~Infeld,
  Proc.\ Roy.\ Soc.\ Lond.\ A {\bf 144}, no. 852, 425 (1934).
  doi:10.1098/rspa.1934.0059


\bibitem{Deffayet:2013tca} 
  C.~Deffayet, A.~E.~Gümrükçüoğlu, S.~Mukohyama and Y.~Wang,
  JHEP {\bf 1404}, 082 (2014)
  doi:10.1007/JHEP04(2014)082
  [arXiv:1312.6690 [hep-th]].

%
\bibitem{Horndeski:1976gi} 
  G.~W.~Horndeski,
  J.\ Math.\ Phys.\  {\bf 17}, 1980 (1976).
  doi:10.1063/1.522837
  
\bibitem{Ostrogradsky:1850fid} 
  M.~Ostrogradsky,
  Mem.\ Acad.\ St.\ Petersbourg {\bf 6}, no. 4, 385 (1850).
;  
  R.~P.~Woodard,
  Scholarpedia {\bf 10}, no. 8, 32243 (2015)
  doi:10.4249/scholarpedia.32243
  [arXiv:1506.02210 [hep-th]].

\bibitem{Langlois:2015cwa} 
  D.~Langlois and K.~Noui,
  JCAP {\bf 1602}, no. 02, 034 (2016)
  doi:10.1088/1475-7516/2016/02/034
  [arXiv:1510.06930 [gr-qc]].
  
\bibitem{Bekenstein:1992pj} 
  J.~D.~Bekenstein,
  Phys.\ Rev.\ D {\bf 48}, 3641 (1993)
  doi:10.1103/PhysRevD.48.3641
  [gr-qc/9211017].

\bibitem{Bettoni:2013diz} 
  D.~Bettoni and S.~Liberati,
  Phys.\ Rev.\ D {\bf 88}, 084020 (2013)
  doi:10.1103/PhysRevD.88.084020
  [arXiv:1306.6724 [gr-qc]].

\bibitem{Zumalacarregui:2013pma} 
  M.~Zumalacárregui and J.~García-Bellido,
  Phys.\ Rev.\ D {\bf 89}, 064046 (2014)
  doi:10.1103/PhysRevD.89.064046
  [arXiv:1308.4685 [gr-qc]].

\bibitem{Jimenez:2016isa} 
  J.~Beltran Jimenez and L.~Heisenberg,
  Phys.\ Lett.\ B {\bf 757}, 405 (2016)
  doi:10.1016/j.physletb.2016.04.017
  [arXiv:1602.03410 [hep-th]].

\bibitem{Kimura:2016rzw} 
  R.~Kimura, A.~Naruko and D.~Yoshida,
  JCAP {\bf 1701}, no. 01, 002 (2017)
  doi:10.1088/1475-7516/2017/01/002
  [arXiv:1608.07066 [gr-qc]].

\bibitem{Domenech:2018vqj} 
  G.~Domènech, S.~Mukohyama, R.~Namba and V.~Papadopoulos,
  Phys.\ Rev.\ D {\bf 98}, no. 6, 064037 (2018)
  doi:10.1103/PhysRevD.98.064037
  [arXiv:1807.06048 [gr-qc]].
  
\bibitem{Ratra:1991bn} 
  B.~Ratra,
  Astrophys.\ J.\  {\bf 391}, L1 (1992).
  doi:10.1086/186384;
%

  V.~Demozzi, V.~Mukhanov and H.~Rubinstein,
  JCAP {\bf 0908}, 025 (2009)
  doi:10.1088/1475-7516/2009/08/025
  [arXiv:0907.1030 [astro-ph.CO]];
  %
  
  S.~Kanno, J.~Soda and M.~a.~Watanabe,
  JCAP {\bf 0912}, 009 (2009)
  doi:10.1088/1475-7516/2009/12/009
  [arXiv:0908.3509 [astro-ph.CO]];
  %
  
  N.~Barnaby, R.~Namba and M.~Peloso,
  Phys.\ Rev.\ D {\bf 85}, 123523 (2012)
  doi:10.1103/PhysRevD.85.123523
  [arXiv:1202.1469 [astro-ph.CO]];
  %
  
  T.~Fujita and S.~Mukohyama,
  JCAP {\bf 1210}, 034 (2012)
  doi:10.1088/1475-7516/2012/10/034
  [arXiv:1205.5031 [astro-ph.CO]];
%

  T.~Fujita and S.~Yokoyama,
  JCAP {\bf 1403}, 013 (2014)
  Erratum: [JCAP {\bf 1405}, E02 (2014)]
  doi:10.1088/1475-7516/2014/03/013, 10.1088/1475-7516/2014/05/E02
  [arXiv:1402.0596 [astro-ph.CO]];
%

  R.~J.~Z.~Ferreira, R.~K.~Jain and M.~S.~Sloth,
  JCAP {\bf 1406}, 053 (2014)
  doi:10.1088/1475-7516/2014/06/053
  [arXiv:1403.5516 [astro-ph.CO]];
%

  H.~Bazrafshan Moghaddam, E.~McDonough, R.~Namba and R.~H.~Brandenberger,
  Class.\ Quant.\ Grav.\  {\bf 35}, no. 10, 105015 (2018)
  doi:10.1088/1361-6382/aaba22
  [arXiv:1707.05820 [astro-ph.CO]].

\bibitem{Neronov:1900zz} 
  A.~Neronov and I.~Vovk,
  Science {\bf 328}, 73 (2010)
  doi:10.1126/science.1184192
  [arXiv:1006.3504 [astro-ph.HE]];
%

  F.~Tavecchio, G.~Ghisellini, L.~Foschini, G.~Bonnoli, G.~Ghirlanda and P.~Coppi,
  Mon.\ Not.\ Roy.\ Astron.\ Soc.\  {\bf 406}, L70 (2010)
  doi:10.1111/j.1745-3933.2010.00884.x
  [arXiv:1004.1329 [astro-ph.CO]];
%

  K.~Dolag, M.~Kachelriess, S.~Ostapchenko and R.~Tomas,
  Astrophys.\ J.\  {\bf 727}, L4 (2011)
  doi:10.1088/2041-8205/727/1/L4
  [arXiv:1009.1782 [astro-ph.HE]];
%

  W.~Essey, S.~Ando and A.~Kusenko,
  Astropart.\ Phys.\  {\bf 35}, 135 (2011)
  doi:10.1016/j.astropartphys.2011.06.010
  [arXiv:1012.5313 [astro-ph.HE]];
%

  A.~M.~Taylor, I.~Vovk and A.~Neronov,
  Astron.\ Astrophys.\  {\bf 529}, A144 (2011)
  doi:10.1051/0004-6361/201116441
  [arXiv:1101.0932 [astro-ph.HE]];
%

  K.~Takahashi, M.~Mori, K.~Ichiki, S.~Inoue and H.~Takami,
  Astrophys.\ J.\  {\bf 771}, L42 (2013)
  doi:10.1088/2041-8205/771/2/L42
  [arXiv:1303.3069 [astro-ph.CO]];
  %
  
  W.~Chen, J.~H.~Buckley and F.~Ferrer,
  Phys.\ Rev.\ Lett.\  {\bf 115}, 211103 (2015)
  doi:10.1103/PhysRevLett.115.211103
  [arXiv:1410.7717 [astro-ph.HE]];
%

  J.~D.~Finke, L.~C.~Reyes, M.~Georganopoulos, K.~Reynolds, M.~Ajello, S.~J.~Fegan and K.~McCann,
  Astrophys.\ J.\  {\bf 814}, no. 1, 20 (2015)
  doi:10.1088/0004-637X/814/1/20
  [arXiv:1510.02485 [astro-ph.HE]].

\bibitem{Durrer:2013pga} 
  R.~Durrer and A.~Neronov,
  Astron.\ Astrophys.\ Rev.\  {\bf 21}, 62 (2013)
  doi:10.1007/s00159-013-0062-7
  [arXiv:1303.7121 [astro-ph.CO]].

\bibitem{Mukohyama:2016npi} 
  S.~Mukohyama,
  Phys.\ Rev.\ D {\bf 94}, no. 12, 121302 (2016)
  doi:10.1103/PhysRevD.94.121302
  [arXiv:1607.07041 [hep-th]].

\bibitem{Heisenberg:2018acv} 
  L.~Heisenberg,
  JCAP {\bf 1810}, 054 (2018)
  doi:10.1088/1475-7516/2018/10/054
  [arXiv:1801.01523 [gr-qc]].

\bibitem{Heisenberg:2014rta} 
  L.~Heisenberg,
  JCAP {\bf 1405}, 015 (2014)
  doi:10.1088/1475-7516/2014/05/015
  [arXiv:1402.7026 [hep-th]].

\bibitem{Hull:2015uwa} 
  M.~Hull, K.~Koyama and G.~Tasinato,
  Phys.\ Rev.\ D {\bf 93}, no. 6, 064012 (2016)
  doi:10.1103/PhysRevD.93.064012
  [arXiv:1510.07029 [hep-th]].


 

\end{thebibliography}
\end{document}